\documentclass[aps,prb,groupedaddress,twocolumn,superscriptaddress]{revtex4}
\usepackage{graphicx}
\usepackage{bm}
\usepackage{epsfig}
\usepackage{ulem}
\usepackage{color}
\usepackage{mathrsfs}
\usepackage{dcolumn}
\usepackage{setspace}
\usepackage{array}
\usepackage{amsmath}
\usepackage{amssymb}
\usepackage{gensymb}
\usepackage{bibentry,natbib}
\usepackage{booktabs}
\begin{document}
\bibliographystyle{unsrt}

\title{Topological Imbert-Fedorov shift in Weyl semimetals}

\author{Qing-Dong Jiang}
\thanks{These authors contributed equally to this work.}
\affiliation{International Center for Quantum Materials, School of
Physics, Peking University, Beijing 100871, P.R. China}
\author{Hua Jiang}
\thanks{These authors contributed equally to this work.}
\affiliation{College of Physics, Optoelectronics and Energy, Soochow
University, Suzhou 215006, P.R. China}
\author{Haiwen Liu}
\affiliation{International Center for Quantum Materials, School of Physics, Peking University, Beijing 100871, P.R. China}
\affiliation{Collaborative Innovation Center of Quantum Matter, Beijing 100871, P.R. China}
\author{Qing-Feng Sun}
\email[]{sunqf@pku.edu.cn}
\affiliation{International Center for Quantum Materials, School of Physics, Peking University, Beijing 100871, P.R. China}
\affiliation{Collaborative Innovation Center of Quantum Matter, Beijing 100871, P.R. China}
\author{X. C. Xie}
\email[]{xcxie@pku.edu.cn}
\affiliation{International Center for Quantum Materials, School of Physics, Peking University, Beijing 100871, P.R. China}
\affiliation{Collaborative Innovation Center of Quantum Matter, Beijing 100871, P.R. China}
\begin{abstract}
The Goos-H\"{a}nchen (GH) shift and the Imbert-Fedorov (IF) shift are optical phenomena which describe the longitudinal and transverse lateral shifts at the reflection interface, respectively.
Here, we report the GH and IF shifts in Weyl semimetals (WSMs)---a promising material harboring low energy Weyl fermions, a massless fermionic cousin of photons.
Our results show that GH shift in WSMs is valley-independent which is analogous to that discovered in a 2D relativistic material---graphene. However, the IF shift has never been explored in non-optical systems, and here we show that it is valley-dependent.
Furthermore, we find that the IF shift actually originates from the topological effect of the system.
Experimentally, the topological IF shift can be utilized to characterize the Weyl semimetals, design valleytronic devices of high efficiency, and measure the Berry curvature.
\end{abstract}
\maketitle
\textit{Introduction.---}Modern quantum physics originates from understanding the
wave-particle duality of all particles, among which  photon is the
first one being discovered with such duality. Classically, the
physics of a beam of light being reflected at an interface is
governed by geometric optics law, where the photons are treated as
classical particles. In contrast, when
considering the wave nature of photons, spatial shifts at the interface appear as longitudinal
shift in the incident plane\cite{goos1949neumessung,bretenaker1992direct,peccianti2006tunable},
or transverse
shift normal to the incident plane\cite{Fedorov,imbert1972calculation,onoda2004hall,bliokh2006conservation,hosten2008observation,yinxiaobolighthall},
which are known as the Goos-H\"{a}nchen (GH) effect and
the Imbert-Fedorov (IF) effect, respectively. Due to
all particles possessing the wave-particle duality, the spatial
shifts are also expected for other particles. For example, GH effect
has been shown to exist in the systems of
electrons\cite{miller1972shifts,fradkin1974spatial,chen2013electronic},
neutrons\cite{de2010observation}, atoms\cite{huang2008goos}, etc. Particularly for the 2D massless Dirac fermions in graphene
systems, the GH shift can be manipulated from positive to negative by
tuning an external electric field\cite{beenakker2009quantum}. However, the
IF effect has not been studied in non-optical systems.

Similar to photons, Weyl particles are also massless. But different
from photons, Weyl particles are spin 1/2 chiral fermions and
described by the Weyl equation.  Recently,  Weyl semimetals (WSMs)
have been proposed  as promising systems embedding Weyl fermions,
generating intensive
interests\cite{wan2011topological,fangchernsemimetal,burkov2011weyl,
luweylpoints,hosur2012charge,jiang2012tunable,yangshenyuan}. In
WSMs, the Weyl nodes always exist in pairs with opposite
chiralities, and each Weyl node corresponds to a valley
index\cite{Nielsen-Ninomiya1981,Nielsen-Ninomiya1983}. Several
candidates are suggested to be WSMs including pyrochlore
irradiates\cite{wan2011topological}, topological insulator and
normal insulator heterostructures\cite{burkov2011weyl}, staggered
flux states in cold atom systems\cite{jiang2012tunable}, and
photonic crystals based on double-gyroid
structures\cite{luweylpoints}. Despite of various theoretical
prediction of WSMs, the experimental realization of WSMs remain a
challenge. This failure cannot be ascribed to the impediment of
material growth technique\cite{Yanagishima,Bramwell}, but to the
scarcity of the method of direct experimental identification of
WSMs\cite{Hosur-Qi}.
Due to the topological properties of the Weyl
fermions, the WSMs may possess exotic wave-packet dynamics, which
indicates a new route to characterizing WSMs and potential
applications in
valleytronics\cite{Rycerzvalley,xiao,makvalley,Gorbachevvalley}.

In this paper, we report the GH effect and IF effect for 3D
Weyl fermions in WSMs.  By
using wave packet method, we derive analytic results for the spatial
shift of the GH effect and IF effect. Our results show that the GH shift
is valley-independent. By contrast, the IF shift is valley-dependent[see Fig. \ref{fig:1}], which give rise to the
valley-dependent anomalous velocities in the system. Due to the IF shift being perpendicular to
the incident plane, it is a generic 3D effect and could never appear
in a 2D material, e.g. the graphene\cite{beenakker2009quantum}. Furthermore, we
demonstrate that the IF shift originates from the topological effect
of the system, namely, the Berry curvature of the system.
Remarkably, the consequence of the valley-dependent anomalous velocity
is significant enough to be detected experimentally. Finally, we
discuss three applications of the valley-dependent IF shift: (i)
effectively characterizing the WSMs; (ii) directly detecting the
Berry curvature; (iii) efficiently inducing valley current with a high
polarization rate.
\begin{figure}[!htb]
\centering
\includegraphics[height=3cm, width=6.cm, angle=0]{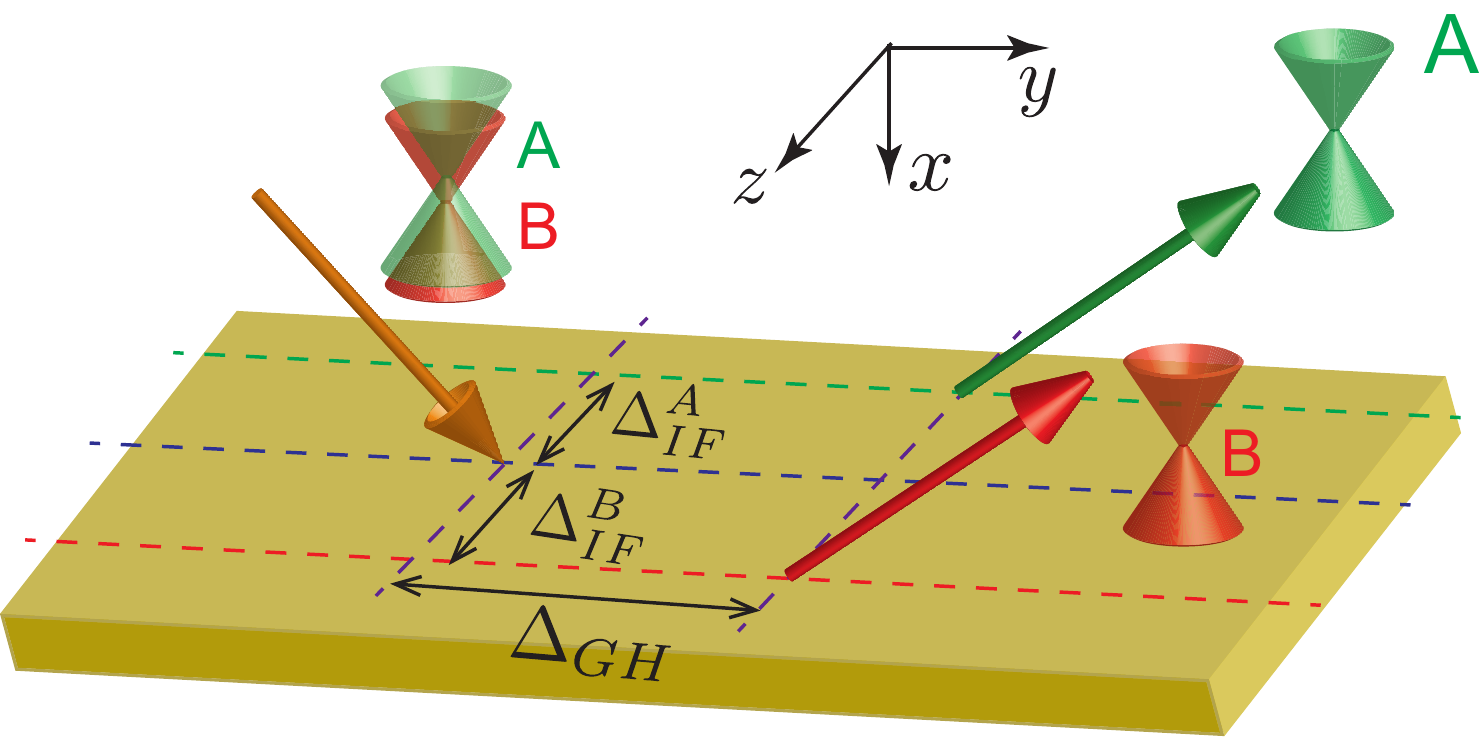}
\caption{Illustration of GH effect and IF effect in WSMs.
The orange arrow represents the incident  wave
packet which include Weyl fermions from two valleys, whereas the
green (red) arrow represents the reflected wave packet of Weyl
fermions only from Valley A (B). $\Delta_{GH}$  denotes the GH
shift, and $\Delta^{A(B)}_{IF}$ stands for the IF shift of valley A
(B).
This figure only shows the positive GH shift
case. \label{fig:1}}
\end{figure}

\textit{Quantum GH and IF effects in WSMs.---}The
Hamiltonian of the WSMs system is
\begin{equation}\label{eq:1}
\mathcal H=\left\{
\begin{array}{ll}
\sum\limits_{i=x,y,z} v_i\,\hat{p}_i\,\sigma_i&(x\leqslant 0)\\
\sum\limits_{i=x,y,z}v_i^\prime\,\hat{p}_i\,\sigma_i+V(x)&(x>0)
\end{array}\right.,
\end{equation}
where $v_i$ ($v_i^\prime$) is velocity parameter for region
$x\leqslant 0$ ($x>0$); $\hat{p}_i$ are momentum
operators, $V(x)$ is potential, and $\sigma_i$ stand for Pauli matrices. To guarantee the model Hamiltonian Hermitian, we let $v_x=v_x^{\prime}$ in this paper. This Hamiltonian
indicates an interface located at $x=0$ in the WSMs[see Fig.
\ref{fig:2}(a)]. We consider a beam of Weyl fermions incident from the
region $x<0$ modeled by a
Gaussian wave packet as $\psi^{in}_g(\bold
r)=\int_{-\infty}^{\infty}\int_{-\infty}^{\infty}d\,k_yd\,k_z\,f(k_y-\bar
k_y)f(k_z-\bar k_z)\,\psi^{in}(\bold k, \bold r)$, where $f(k_s-\bar
k_s)=(\sqrt{2\pi}\Delta_{k_s})^{-1}~e^{-(k_s-\overline
k_s)^2/2\Delta^2_{k_s}}$ are Gaussian distribution functions of
width $\Delta_{k_s}$ peaked at the mean wave vector $(\bar k_x,\bar k_y,\bar k_z)$ with
$s=y\,,z$. Note that none of our results depend on the shape of the wave packet. Here, $\psi^{in}(\bold k, \bold r)$ is the incident wave
function, which is a solution of Weyl equation, i.e., $\mathcal
H\psi^{in}=E\psi^{in}$ for region $x<0$:
\begin{eqnarray}\label{eq:2}
\psi^{in}(\bold k,\bold r)=\frac{1}{\sqrt{1+\eta^2}}\left(\begin{array}{cc}
e^{-i\alpha/2}\\ \eta\,e^{i\alpha/2}
\end{array}\right)\,e^{ik_x x+i k_y y+i k_z z},
\end{eqnarray}
where $ k_x=\sqrt{E^2- (\hbar v_y  k_y)^2-(\hbar v_z k_z)^2}/\hbar
v_x$, $\alpha=\tan^{-1}(v_y\,k_y/v_x\,k_x)$, and
$\eta=\sqrt{\frac{E-\hbar v_{z}k_z}{E+\hbar v_{z}k_z}}$.
Analogously, the reflected wave packet can be written as
$\psi^{re}_g(\bold
r)=\int_{-\infty}^{\infty}\int_{-\infty}^{\infty}d\,k_yd\,k_z\,
f(k_y-\bar k_y)f(k_z-\bar k_z)\,\psi^{re}(\bold k, \bold r)$, where
$\psi^{re}$ is the reflected wave function. $\psi^{re}$ can be
obtained from the incident wave Eq.(\ref{eq:2}) by the substitution
$k_x\mapsto-k_x$, $\alpha\mapsto\pi-\alpha$ and multiplication with
the reflection amplitude $r=\vert{r}\vert e^{i \phi_r}$.
The integrals of $\psi_g^{in}$ and $\psi_g^{re}$ give the center of the wave packets, and therefore we can obtain the spatial shifts in  $y$, $z$ directions\cite{footnote3}:
\begin{eqnarray}\label{eq:4}
\left\{
\begin{array}{cc}
\Delta^{y}_{\pm}=-\frac{\partial}{\partial\,k_y}\phi_r(\bar k_y,\bar k_z)\mp\frac{\partial}{\partial\,k_y}\alpha(\bar k_y, \bar k_z)\\
\Delta^{z}_{\pm}=-\frac{\partial}{\partial\,k_z}\phi_r(\bar k_y,
\bar k_z)\mp\frac{\partial}{\partial\,k_z}\alpha(\bar k_y, \bar k_z)
\end{array}\right..
\end{eqnarray}
The spatial shifts for Weyl fermions are defined as
the average shifts of the two spinor components:
$\Delta^{y(z)}=(\Delta_{+}^{y(z)}+\eta^2\Delta_{-}^{y(z)})/(1+\eta^2)$.
In Eq.(3), $\alpha(\bar k_y,\bar k_z)$ represents the incident angle, and $\phi_r$ is the phase of the reflection coefficient, which
can be obtained by matching the wave function at $x=0$. If the incident wave packet is confined in $x$-$y$ plane, the
in-plane shift $\Delta^y$ and out-of-plane shift $\Delta^z$ of wave packet corresponds to the GH shift $\Delta_{GH}$ and IF shift $\Delta_{IF}$:
\begin{eqnarray}
\Delta_{GH}&=&\frac{v_y^\prime}{v_x^\prime}\,\,\frac{
(1+\beta\,\sin^2\bar{\alpha}-\frac{V}{E})}{\beta\,\sin\bar{\alpha}\,
\cos\bar{\alpha}\, \kappa},\label{eq:5}\\
 \Delta_{IF}&=&-C\,\vert\frac{\hbar v_x
v_z}{v_y}\vert\frac{1}{E\,
\tan\theta}\left[\frac{1-\frac{E}{V}(1-\gamma)}{1-\frac{E}{V}(1-\beta)}\right],\label{eq:6}
\end{eqnarray}
where $\kappa=\sqrt{(\hbar v_y^\prime k_y)^2+(\hbar
v_z^\prime k_z)^2-(E-V)^2}/\hbar v_x^\prime$, $\beta=v_y^\prime/v_y$, $\gamma=v_z^\prime/v_z$,
$\bar{\alpha}=\tan^{-1}\,(v_y\,\bar k_y/v_x\,\bar k_x)$, and
$\theta=\tan^{-1}\,(\bar k_y/\bar k_x)$. $C\equiv sgn[v_x\,v_y\,v_z]$ is the chirality of
the valley in WSMs.

\begin{figure}[!htb]
\centering
\includegraphics[height=6.cm, width=8.cm, angle=0]{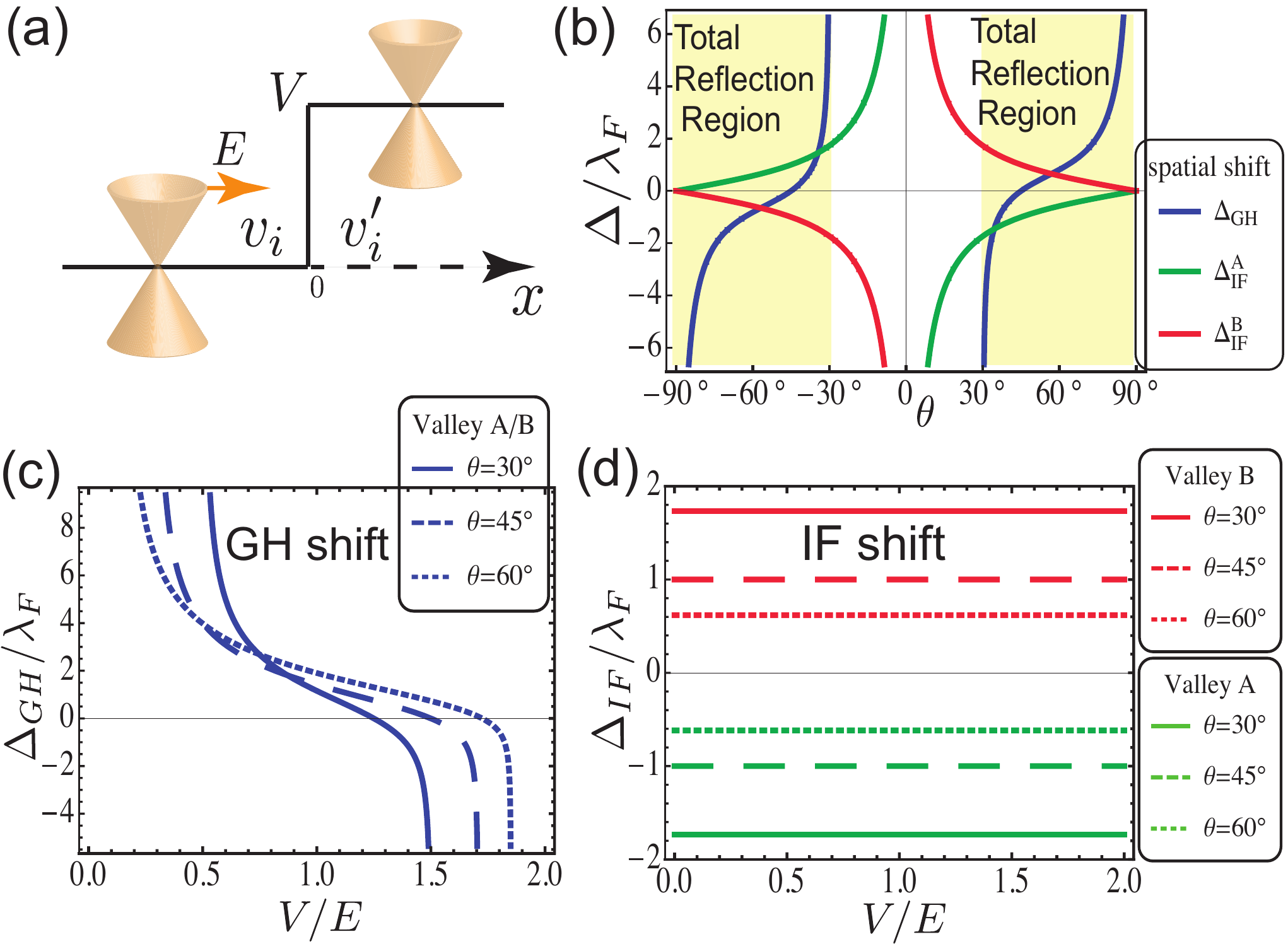}
\caption{
(a) Schematic of p-n junction with an interface at $x=0$. $v_i$
($v_i^\prime$) is the velocity in the left (right) side of the
interface. $E$ is the Fermi energy, and $V$ is the potential
difference of the junction. (b)  GH shift $\Delta_{GH}$ and IF shift
$\Delta^{A(B)}_{IF}$ of valley A (B) versus incident angle $\theta$.
Total reflection happens at $\theta\geqslant30\degree$ (the yellow
color shadow region).  $\lambda_F=h\,v_i/E$ stands for the wave
length of Weyl fermions. (c) Potential dependence of the GH shift
for incident angle $\theta=30\degree$ (solid), $45\degree$ (dashed),
and $60\degree$ (dotted). (d) Potential dependence of the IF shift
for incident angle $\theta=30\degree$ (solid), $45\degree$ (dashed),
and $60\degree$ (dotted). \label{fig:2}}
\end{figure}

We take the incident Fermi energy $E=100~{\rm
meV}$, potential $V=150~{\rm meV}$, and velocities
$v_i=v_i^\prime=10^6~{\rm m/s}$, where $i=x\,,y\,,z$. In this case,
$\beta=\gamma=1$. Note that the two valleys (A/B) of a WSMs
have opposite chiralities. Fig. \ref{fig:2}(b) shows the spatial
shifts versus the incident angle $\theta$, where both the GH and IF
shifts are odd functions of incident angle, which is consistent with
symmetry analysis.
Fig. \ref{fig:2}(c) shows the valley independence and potential dependence of GH
shift $\Delta_{GH}$, which can be tuned from
positive to negative by external field $V$. This feature is analogous to the GH shift in graphene\cite{beenakker2009quantum}.
Fig. \ref{fig:2}(d) illustrates that the IF shift is independent of
potential, but depends on valley index.  The IF shift
can be utilized to manipulate the valley degree of freedom.
\begin{figure}
\centering
\includegraphics[height=7cm, width=8.cm, angle=0]{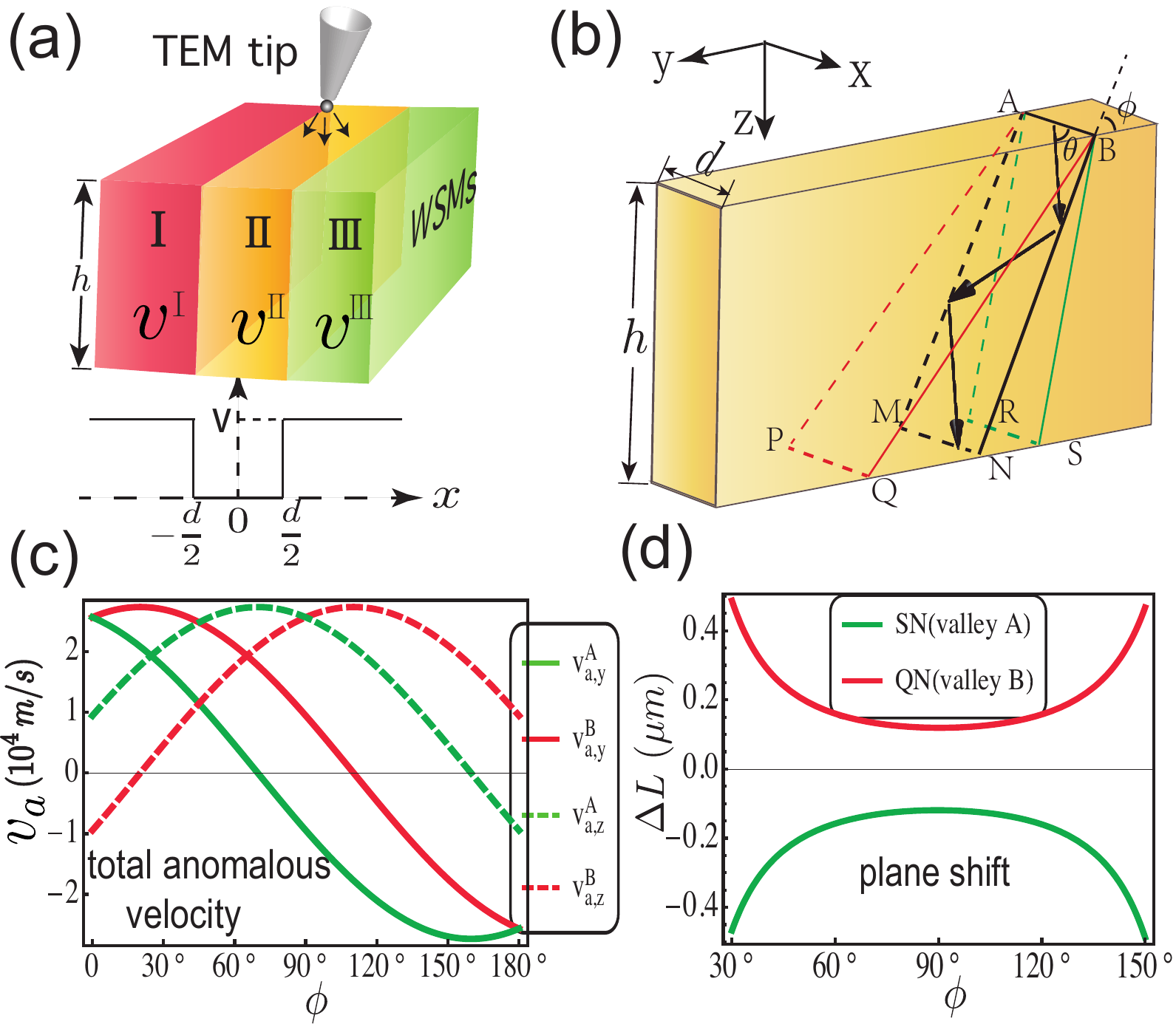}
\caption{
(a) Schematic of the valley splitter for Weyl fermions.
Region \uppercase\expandafter{\romannumeral1},
\uppercase\expandafter{\romannumeral2}, and
\uppercase\expandafter{\romannumeral3} are three WSM
layers with different velocities  $\bold
v^{\uppercase\expandafter{\romannumeral1}}$, $\bold
v^{\uppercase\expandafter{\romannumeral2}}$, and $\bold
v^{\uppercase\expandafter{\romannumeral3}}$.
(b) Illustration of wave packet trajectory (black arrow) of
Weyl fermions in region \uppercase\expandafter{\romannumeral2}. The
electrons are injected by TEM tip with incident polar angle $\theta$
and azimuthal angle $\phi$.
The parameter $\theta$ is fixed $50\degree$ in our calculation. In
ordinary case, the trajectory of wave packet stays in ABMN plane. However, due to the valley-dependent IF shift, the
trajectory of Weyl fermions from valley A (B) would shift to plane
ABRS (ABPQ).
(c) $\phi$ dependence of the total
anomalous velocities. (d) The propagating plane shift for two
valleys. After considering IF shift, SN (QN) is the final
position shift for valley A (B).
\label{fig:3}}
\end{figure}

\textit{Topological origin of the IF effect.---}Based on the semiclassical dynamics of wave packet, we show that the
IF shift is closely related to the Berry curvature of the system.
Let us assume the velocity $v_i=v_i^\prime$ ($i=x,\,y\,,z$)
in WSMs system. In this case, $\beta=\gamma=1$, and Eq.(\ref{eq:6})
reduces to $\Delta_{IF}=-(\frac{\hbar v_x\, v_{z}}{v_y})\frac{1}{E\,
\tan\theta}$. Commonly, the Weyl node can be regarded as a magnetic monopole
in $k$-space\cite{fang}, and thereby generates an effective magnetic
field in $k$-space.  The
Berry curvature of Hamiltonian Eq.(\ref{eq:1})  in region $(x\leq0)$
is $\bold\Omega^{\pm}=\mp\frac{\hbar^3 v_x v_y v_z \bold k}{2\,E^3}$
for conduction band and valence band\cite{xiao2010berry},
respectively.  The semiclassical equation of motion (EOM) \cite{xiao2010berry,chang1996berry} of wave packet is
$\frac{d\,\bold r}{d\,t}=\frac{\partial E(\bold k)}{\hbar\,\partial\bold k}-\frac{d\,\bold k}{d\,t}\times\bold\Omega$,
where $\bold r$ and $\bold k$ are the center positions of the wave
packet in phase space.  We assume that the incident wave packet locates
in the conduction band, and consider the incident
wave packet in x-y plane with $k_z=0$.  The variation of $k_x$ by potential V and nonzero Berry curvature $\Omega_y$ leads to the IF shift $\Delta_{IF}$ in $z$-direction:
\begin{equation}\label{eq:8}
\Delta_{IF}=-\int_{k_x}^{-k_x}
d\,k_x\,\Omega_y=-(\frac{\hbar v_x\, v_{z}}{v_y})\frac{1}{E\,
\tan\theta},
\end{equation}
where $\tan\theta=k_y/k_x$. Remarkably, Eq.(\ref{eq:8}) and
Eq.(\ref{eq:6}) completely coincide in the case of $\beta=\gamma$.
This coincidence does not depend on linear dispersion of the system\cite{footnote3}.
The consistency between
Berry curvature calculations and wave packet results strongly
support that the IF shift is mainly a topological effect.
The IF shift in WSMs is quite different from that in optical systems.  Particularly, $\Delta_{IF}$ reaches the maximum in WSMs [see Fig. 2(b)] comparing the zero value in optical systems at $\theta= 0\degree$\cite{onoda2004hall,hosten2008observation}.  This is because the conservation of $k_y,k_z$ guarantee the Weyl Fermions stay in the same valley during the reflection processes in WSMs. In contrast, the polarization  of the photons change during the reflection processes, which will severely influence $\Delta_{IF}$.
However,the semiclassical equation cannot be generally applied to all
cases of reflection process because of the breakdown of adiabatic approximation for some systems.

\textit{Anomalous velocities induced by IF effect.---}We consider a well collimated beam of Weyl fermions
propagating in the middle layer (Region
\uppercase\expandafter{\romannumeral2}) of a sandwich structure, which is constructed by three
layers of WSMs[see Fig. \ref{fig:3}(a)]. The applied electric
potential profile is shown below the sandwich structure. The height of the structure is $h$ and the width of
the region \uppercase\expandafter{\romannumeral2} is $d$. There are
both GH and IF shifts at the two interfaces ($x=\pm\frac{d}{2}$).
In order to observe the valley-dependent IF shift, the
mirror symmetry about $x=0$ plane need to be broken\cite{footnote3}. Thus, we consider that the z direction velocities in the three regions are different,
with $v_z^{\rm\uppercase\expandafter{\romannumeral1}}=v_L$,
$v_z^{\rm\uppercase\expandafter{\romannumeral2}}=v$, and
$v_z^{\rm\uppercase\expandafter{\romannumeral3}}=v_R$;
whereas the x and y directions velocities are still identical $v_{x/y}^{\rm
\uppercase\expandafter{\romannumeral1}}=v_{x/y}^{\rm\uppercase\expandafter{\romannumeral2}}=
v_{x/y}^{\rm\uppercase\expandafter{\romannumeral3}}=v$.
Without considering the spatial shifts at the interfaces,
the normal velocities in region
\uppercase\expandafter{\romannumeral2} in $y$ and $z$ directions are
$v_{n,y}=v\,\sin\theta\,\cos\phi$ and $v_{n,z}=v\,
\sin\theta\,\sin\phi$, where angles $\theta$ and $\phi$ characterize
the incident direction of the wave packet [see Fig. 3(b)].
We denote the GH and IF shifts at the left
(right) interface as $\Delta^{L(R)}_{GH}$ and $\Delta^{L(R)}_{IF}$,
respectively. During multiple reflections in region
\uppercase\expandafter{\romannumeral2}, the GH and IF shifts
are accumulated, and induce average anomalous velocities:
$v_{a,y}=\left[(\Delta^{L}_{GH}+\Delta^{R}_{GH}) \cos\phi+
(\Delta^{L}_{IF}+\Delta^{R}_{IF}) \sin\phi\right]/(2\Delta t)$ and
$v_{a,z}=[(\Delta^{L}_{GH}+\Delta^{R}_{GH}) \sin\phi+
(\Delta^{L}_{IF}+\Delta^{R}_{IF}) \cos\phi]/(2 \Delta t)$. $\Delta
t= d/(v \,\cos\theta)$ represents the propagating time between two
subsequent reflections. Therefore, the normal and anomalous velocity result in the total velocity: $v_{t,y(z)}=v_{n,y(z)}+v_{a,y(z)}$.

We set the parameters
$E=100 ~{\rm meV}$, $V=150~{\rm
meV}$, $h=10{\rm\mu m}$, $d=50~{\rm nm}$,
$
v^{\uppercase\expandafter{\romannumeral1}}_z=v_L=1.2\times10^6~{\rm
m/s}$, $ v^{\uppercase\expandafter{\romannumeral2}}_z=v=10^6~{\rm
m/s}$, $
v^{\uppercase\expandafter{\romannumeral3}}_z=v_R=0.8\times10^6~{\rm
m/s}$, and velocities in other directions are all set as $10^6~{\rm
m/s}$\cite{footnote2}.
Fig. \ref{fig:3}(c) shows $\phi$ dependence of the total anomalous velocities induced by the spatial shifts.
The anomalous velocities are valley-dependent, which implies that the total
velocities also depend on valley index. Eventually, the different velocities of the two valleys lead to macroscopic separation in real space, which is experimentally detectable.
Fig. \ref{fig:3}(d) shows the opposite position shift at the bottom of region \uppercase\expandafter{\romannumeral2} for valley A  and valley B. GH shift cannot induce the
position shift, because GH shift always lie in the propagating
plane. In contrast, IF shift can induce the position shifts (${\rm
\mu m}$ order) for two valleys ($\bold{SN}$
and $\bold{QN}$)\cite{footnote3}.

\begin{figure}
\centering
\includegraphics[height=7.0cm, width=8.cm, angle=0]{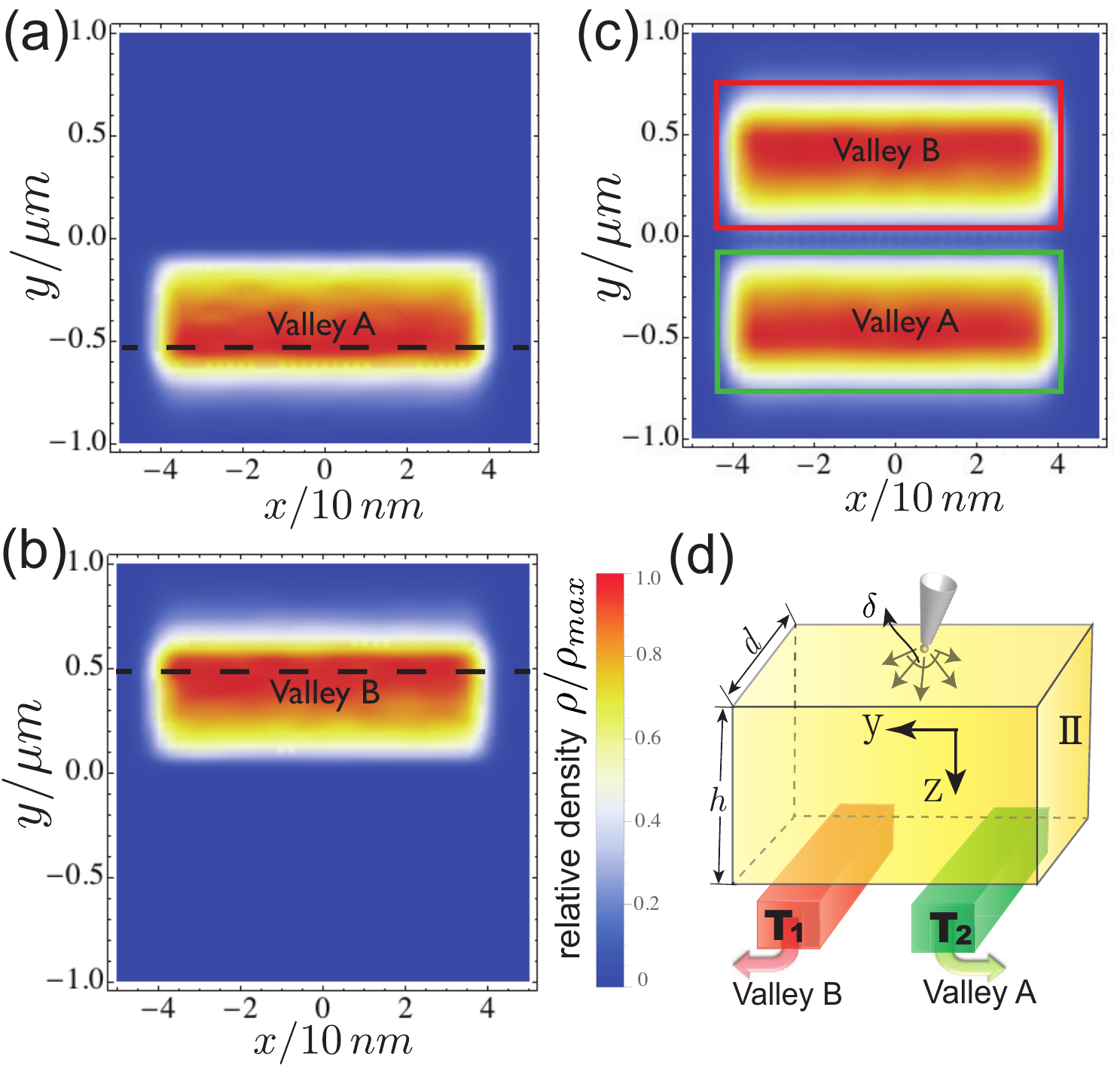}
\caption{Valley density distribution shift caused by IF effect.
(a), (b) and (c) show the relative density distribution of
valley A, valley B, and both, respectively, after considering GH and IF effect.
The color bar on the right represents the relative density $\rho/\rho_{max}$ of
Weyl fermions. Here, $\rho$ ($\rho_{max}$) represents the absolute (maximum) density of Weyl fermions at the bottom.
The dashed lines indicate the location of the maximum
intensity of density. Red
(Green) box in Fig. (c) guides eyes to the region where only valley B (A) exist
(pure valley polarization).
(d) Proposed setup for
generating valley current. The devices $\bold T_1$ (red) and $\bold
T_2$ (green) are two terminals, which can extracted valley current out. \label{fig:4}}
\end{figure}

\textit{Identification of WSMs.---}The experimental verification of
WSMs has not been justified mainly due to the scant of efficient
detection method\cite{Aji2012,Spivak2013,Parameswaran}. The direct
ARPES measurements of energy dispersions are currently scant due to
the constraint of magnetic properties of WSMs. Here, we suggest that
the IF effect can be used as an experimental identification of WSMs.
It has been shown that the topological IF shift splits the incident
wave packets into opposite directions respect to the valley index.
Thus, the splitting of wave packets on the bottom of region
${\rm\uppercase\expandafter{\romannumeral2}}$ in Fig. \ref{fig:3}(a)
can serve as a hallmark of WSMs. Furthermore, even considering an
incident wave packet with finite angle range, this exotic splitting
can also exist.

\textit{Pure valley polarization.---}
Let us consider an TEM injector with incident angle range $\delta$
($\theta\in[\theta_{c}-\delta/2, ~\theta_{c}+\delta/2]$, and
$\phi\in[\phi_c-\delta/2, \phi_c+\delta/2]$) at the top of Region
\uppercase\expandafter{\romannumeral2}, and study the valley density distribution at the bottom [see Fig. \ref{fig:4}(d)].
We calculate the valley density distribution with parameters $E=100~{\rm
meV}$, $V=150~{\rm meV}$, $h=10~{\rm \mu m}$, $d=80~{\rm nm}$,
$\delta=3\degree$, $\theta_c=32.5\degree$, and $\phi_c=90\degree$.
Without considering anomalous velocities, the
density distributions for valley A and B are maximized at the center of y direction, and thus not distinguishable\cite{footnote3}. In contrast, after
considering the anomalous velocities, Fig. \ref{fig:4}(a), (b) and (c)
show the density distributions for valley A, valley B, and valley A \& B,
respectively.
The IF effect induces opposite shifts in y-direction for valley A and valley B.
Since the Weyl fermions from valleys A and B are well separated in space
of micrometer order, pure valley current can be generated in the
green (red) regions[see Fig. \ref{fig:4}(c)]. This schematic set-up in Fig. \ref{fig:4}(d)
can be utilized to generate pure valley current.

\textit{Detection of Berry curvature.---}The Berry curvature--- a
gauge invariant quantity---should be detectable in
experiment\cite{xiao2010berry}. However, up to date,  there lacks an
experimental feasible method to measure the Berry curvature in real
materials. The topological IF effect provides a new way to measure
the Berry curvature. For a system with inversion symmetry, the Berry
curvature is an even function of wave vector $\bold k$, i.e., $\bold
\Omega(\bold k)=\bold \Omega(-\bold k)$\cite{xiao2010berry}, which
is usually satisfied in
WSMs\cite{wan2011topological,fangchernsemimetal,burkov2011weyl}.
Considering a wave packet propagating in $x$-$y$ plane with $k_{z} =
0$, the IF shift can be expressed as $\Delta_{IF}=2
\int_{0}^{k_x}dk_x\,\Omega_y(\bold k)$, or
\begin{equation}\label{eq:9}
\Omega_y(\bold k)=\frac{1}{2}\frac{\partial\,\Delta_{IF}(\bold k)}{\partial\,k_x}.
\end{equation}
To measure the Berry curvature $\Omega_y$, one need to collect the IF shift in $z$ direction
$\Delta_{IF}(E, \theta)$ as a function of energy $E$ and incidence
angle $\theta$. The $\Delta_{IF} (E, \theta)$ can be transformed
into $\Delta_{IF} (k_{x}, k_{y})$, and its derivative gives out
the Berry curvature $\Omega_y(\bold k)$.  In the same way,
the IF shift in other directions can be used to obtain
$\Omega_x(\bold k)$ and $\Omega_z(\bold k)$.


\textit{Summary.---}We obtained analytical expressions of the GH shift and IF shift at the interface of WSMs. We demonstrate that the IF shift is valley-dependent, and can be attributed to the topological nature of the system. This IF shift  can lead to valley-dependent anomalous velocity, which is experimentally detectable.  Finally,  we discuss three applications of the topological IF shift including characterization of WSMs, fabrication of high efficient valleytronic devices, and detection the Berry curvature.

\textit{Acknowledgments.---}This work was financially supported by
NBRP of China (2012CB921303, 2012CB821402, and 2014CB920901) and NSF-China under
Grants Nos. 11274364, and 91221302, and 11374219.\\

\hspace{3mm}

\clearpage
\begin{widetext}
\appendix
\section{Supplementary Information for ``Topological Imbert-Fedorov shift in Weyl semimetals"}

\begin{center}
Qing-Dong Jiang, Hua Jiang, Haiwen Liu, Qing-feng Sun, and X. C. Xie
\end{center}
\maketitle

\begin{center}
\textbf{${\rm\uppercase\expandafter{\romannumeral1}}$. Detail derivation of the GH and IF shifts.}
\end{center}

To calculate the spatial shifts of the wave packet in $y$,
$z$ directions, one need to know the central positions of the
incident and the reflected wave packets at the interface ($x=0$). By
expanding the phases $\alpha(k_y, k_z)$ and $\phi_r(k_y,k_z)$ to the
first order around $(\bar k_y, \bar k_z)$, the integrals of
$\psi_g^{in}$ and $\psi_g^{re}$ give that $\psi^{in}_{g\pm}\propto
e^{-(y\mp\frac{1}{2}\frac{\partial\alpha}{\partial
k_y})^2\Delta^2_{k_y}/2}\,\,\,e^{-(z\mp\frac{1}{2}\frac{\partial\alpha}{\partial
k_z})^2\Delta^2_{k_z}/2}$ and $\psi^{re}_{g\pm}\propto
e^{-(y+\frac{\partial\phi_r}{\partial
k_y}\pm\frac{1}{2}\frac{\partial\alpha}{\partial
k_y})^2\Delta^2_{k_y}/2}\,\,\,e^{-(z+\frac{\partial\phi_r}{\partial
k_z}\pm \frac{1}{2}\frac{\partial\alpha}{\partial
k_z})^2\Delta^2_{k_z}/2}$, where the $\pm$ subscript corresponds to
the first and the second component of the spinor, respectively. Then one can identify the center (maximum) of the incident and reflected wave packets in real space. 
For the incident wave packet, the two spinor components are centered at $(\bar y^{in}_{\pm}, \bar z^{in}_{\pm})$, where
\begin{eqnarray}\label{eq:s1}
\left\{
\begin{array}{cc}
\overline y_{\pm}^{in}=\left[\pm\frac{1}{2}\frac{\partial}{\partial\,k_y}\alpha(k_y,k_z)\right]_{k_y=\bar k_y,k_z=\bar k_z}\\
\overline z_{\pm}^{in}=\left[\pm\frac{1}{2}\frac{\partial}{\partial\,k_z}\alpha(k_y,k_z)\right]_{k_y=\bar k_y,k_z=\bar k_z}
\end{array}\right..
\end{eqnarray}
For the reflected wave packet, the two spinor components are centered at $(\bar y^{re}_{\pm}, \bar z^{re}_{\pm})$, where
\begin{eqnarray}\label{eq:s2}
\left\{
\begin{array}{cc}
\overline y_{\pm}^{re}=\left[-\frac{\partial}{\partial\,k_y}\phi_r(k_y,k_z)\mp\frac{1}{2}\frac{\partial}{\partial\,k_y}\alpha(k_y,k_z)\right]_{k_y=\bar k_y,k_z=\bar k_z}\\
\overline z_{\pm}^{re}=\left[-\frac{\partial}{\partial\,k_z}\phi_r(k_y,k_z)\mp\frac{1}{2}\frac{\partial}{\partial\,k_z}\alpha(k_y,k_z)\right]_{k_y=\bar k_y,k_z=\bar k_z}
\end{array}\right..
\end{eqnarray}
Considering above results, we can obtain the spatial shifts for two spinor components in $y$, $z$ directions, which are
\begin{eqnarray}\label{eq:s3}
\left\{
\begin{array}{cc}
\Delta^{y}_{\pm}=\overline y_{\pm}^{re}-\overline y_{\pm}^{in}=-\frac{\partial}{\partial\,k_y}\phi_r(\bar k_y,\bar k_z)\mp\frac{\partial}{\partial\,k_y}\alpha(\bar k_y, \bar k_z)\\
\Delta^{z}_{\pm}=\overline z_{\pm}^{re}-\overline z_{\pm}^{in}=-\frac{\partial}{\partial\,k_z}\phi_r(\bar k_y, \bar k_z)\mp\frac{\partial}{\partial\,k_z}\alpha(\bar k_y, \bar k_z)
\end{array}\right.
\end{eqnarray}

As is shown in Eq. (\ref{eq:s3}), in order to calculate the spatial shift at the interface (x=0), we need to
know the reflection phase $\phi_r(\bar k_y, \bar k_z)$ and $\alpha(\bar k_y, \bar k_z)$.  $\alpha(\bar k_y, \bar k_z)$ represents the incident angle of the wave packet, and can be easily obtained once we know the incident wave vector. Next, we elaborate on deriving the reflection phase $\phi_r$. We consider the total reflection
case, i.e., the reflection probability $\vert r\vert^2=1$. Thus, the
wave function must be evanescent in the region $x>0$[Fig.
2(a)].  We can calculate the reflection coefficient by
matching the wave function $\psi^{in}(\bold k, \bold
r)+\psi^{re}(\bold k, \bold r)$ at $x=0$  to the evanescent wave.
The continuity of wave function gives the reflection amplitude
$r=e^{i\,\phi_r}$, where $\phi_r=2 \,\tan^{-1}(\frac{\eta\,
\cos\alpha}{-\eta \,\sin\alpha+\xi})-\alpha-\pi/2$. In this
expression,  $\xi=\frac{\hbar(\kappa+k_y)}{E-V+\hbar v_z^\prime
k_z}$, where $\kappa=\sqrt{(\hbar v_y^\prime k_y)^2+(\hbar
v_z^\prime k_z)^2-(E-V)^2}/\hbar v_x^\prime$. Substitute the
expression of $\alpha$ and $\phi_r$ into Eq.(\ref{eq:s3}), and  we
can obtain the average shifts in $y$, $z$ directions.

\vspace{5mm}

\begin{center}
\textbf{${\rm\uppercase\expandafter{\romannumeral2}}$. IF shift for nonlinear energy dispersion.}
\end{center}

In the main text, we have shown that the Berry curvature is the origin of the IF shift for WSMs, which possess linear energy band dispersions. Here, we show that the Berry curvature is still the origin of the IF shift for another system without linear energy dispersions. To clarify this idea, we consider a system with a model Hamiltonian
\begin{equation}\label{eq:s4}
\mathcal H^\prime=\left(\begin{array}{cc}
A k_y^2 & B (k_x-i\,k_z)\\
B(k_x+i\,k_z) & -A k_y^2
\end{array}\right),
\end{equation}
where $A$ and $B$ are two parameters. The eigenenergy of the Hamiltonian is $E=\pm\sqrt{B^2(k_x^2+k_z^2)+A^2k_y^4}$. Thus this Hamiltonian obviously does not have 3D linearized energy dispersions.
One can apply a gate voltage $V$ to this system at the region $x\geqslant0$; therefore, an interface appears at $x=0$.

We now consider that a wave packet incidents on the interface from the region $x<0$. To calculate the IF shift at the interface, we follow the same procedure in the main text. Analogously, we first construct the incident wave packet in Gaussian profile with center momentum $(\bar k_x,\bar k_y,\bar k_z)$. Still we  assume the wave packet as $\psi^{in}_g(\bold r)=\int_{-\infty}^{\infty}\int_{-\infty}^{\infty}d\,k_yd\,k_z\,f(k_y-\bar k_y)\,f(k_z-\bar k_z)\,\psi^{in}(\bold k, \bold r)$, where $f(k_s-\bar k_s)$ has the same definition as in the main text. Similarly, the reflected wave packet can be constructed as $\psi^{re}_g(\bold r)=\int_{-\infty}^{\infty}\int_{-\infty}^{\infty}d\,k_yd\,k_z\,f(k_y-\bar k_y)\,f(k_z-\bar k_z)\,\psi^{re}(\bold k, \bold r)$. However, since the Hamiltonian $\mathcal H^\prime$ here is different from that in the main text, one need to solve the wave functions $\psi^{in}(\bold k, \bold r)$ and $\psi^{re}(\bold k, \bold r)$ for $\mathcal H^\prime$. The Hamiltonian equation is $\mathcal H^\prime \psi=E\psi$, where $\psi=(\psi_1,\psi_2)^\mathrm{ T }$ is a two components spinor. Solve this Hamiltonian equation and use continuity condition of the wave function at $x=0$, then we can get the solutions:
\begin{eqnarray}\label{eq:s5}
\psi^{in}(\bold k,\bold r)=\frac{1}{\sqrt{1+\zeta^2}}\left(\begin{array}{cc}
e^{-i\alpha (k_y,k_z)/2}\\ \zeta\,e^{i\alpha (k_y,k_z)/2}
\end{array}\right)\,e^{ik_x x+i k_y y+i k_z z},
\end{eqnarray}
and
\begin{eqnarray}\label{eq:s6}
\psi^{re}(\bold k, \bold r)=\frac{r}{\sqrt{1+\zeta^2}}\left(\begin{array}{cc}
-i\,e^{i\alpha (k_y,k_z)/2}\\ i\,\zeta\,e^{-i\alpha (k_y,k_z)/2}
\end{array}\right)\,e^{-ik_x x+i k_y y+i k_z z}.
\end{eqnarray}
In the above expressions, $ k_x=\sqrt{E^2- (A k_y^2)^2-(B k_z)^2}/B$, $\alpha=tan^{-1}(k_z/k_x)$, $\zeta=\sqrt{\frac{E-\alpha k_y^2}{E+\alpha k_y^2}}$, $r=\vert{r}\vert e^{i \phi_r}$ is the reflection amplitude, and $\phi_r=2\, tan^{-1}(\frac{\zeta cos\,\theta}{-\zeta \, sin\,\theta+\xi^\prime})$ with $\xi^\prime=\frac{B(\kappa+k_z)}{E-V+A k_y^2}$ and $\kappa=\sqrt{(E-V)^2- (A k_y^2)^2-(B k_z)^2}/B$.
Substitute $\phi_r$ and $\alpha$ into Eq.(\ref{eq:s3}), and one can obtain the spatial shifts in $y$ and $z$-directions. If the incident wave packet is confined in $x$-$y$ plane, then the out-of-plane spatial shift in $z$-direction corresponds to the IF shift and reads:
\begin{eqnarray}\label{eq:s7}
\Delta_{IF}=\frac{B^2}{A}\frac{k_x}{ E\,k_y^2}.
\end{eqnarray}

Alternatively, one can also obtain the IF shift $\Delta_{IF}$ from semiclassical equation for wave packet (Eq.(7) in the main text).  In order to apply the semiclassical equation, one need  calculate the Berry curvature $\bold \Omega^{\pm}$ for the Hamiltonian $\mathcal H^\prime$, where the superscript $\pm$ corresponds to the conductance band and valence band, respectively.  The Berry curvature of this system is $\bold \Omega^{\pm}=(\mp \frac{A B^2}{2 E^3}k_x \,k_y,~ \mp \frac{AB^2}{2 E^3}k_y^2,~\mp \frac{AB^2}{2 E^3}k_z \,k_y)$\cite{xiao2010berry}. Substitute the Berry curvature expression into the Eq.(11) in the main text,  and one can obtain the analytic result of the IF shift $\Delta_{IF}=\frac{B^2}{A}\frac{k_x}{ E\,k_y^2}$, showing the  consistence with Eq.(\ref{eq:s7}).  The fact that the IF shift $\Delta_{IF}$ can be obtained from Eq.(7) in the main text provides another convincing argument that IF shift arises from the Berry curvature of the system.

\vspace{5mm}

\begin{center}
\textbf{${\rm\uppercase\expandafter{\romannumeral3}}$. Calculations of anomalous velocities and position shifts for two valleys.}
\end{center}

\begin{figure*}[!htb]
\centering
\includegraphics[height=6cm, width=16.cm, angle=0]{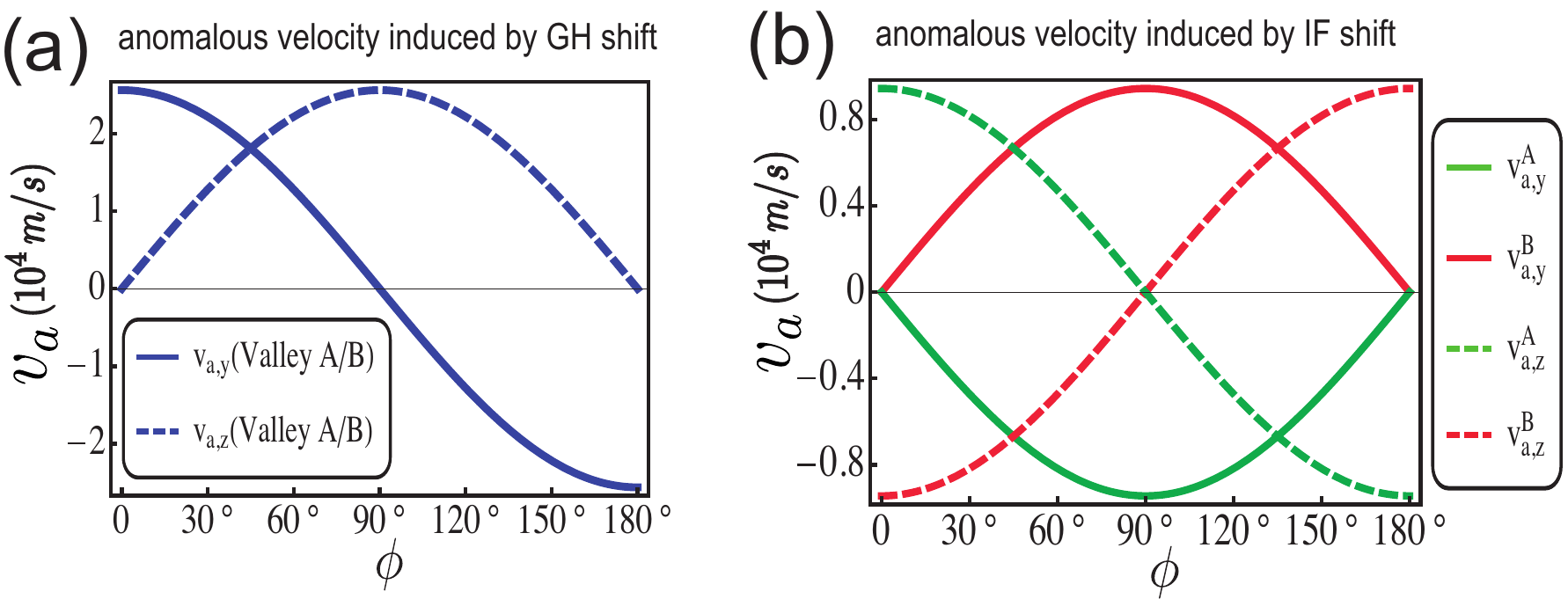}
\caption{(a) The anomalous velocities induced by the GH shift are valley-independent. (b)The anomalous velocities induced by the IF shift are valley-independent.  \label{fig:S1}}
\end{figure*}

In the main text, we have demonstrated that the GH shift does not depend on valley index, whereas the IF shift depends on valley index. Although both of the GH shift and IF shift can bring anomalous velocities to the normal velocities, only the anomalous velocity induced by IF shift is valley-dependent. There are
both GH and IF shifts at the left interface ($x=-\frac{d}{2}$) and
the right interface ($x=\frac{d}{2}$). If the system has the mirror
symmetry about the $x=0$ plane, the GH shifts are in the same
direction at the left and right interfaces. But IF shifts are in
opposite direction at these two interfaces, leading to the cancellation of the IF shifts
after two subsequent reflections.
Therefore, in order to observe the valley-dependent IF shift, the
mirror symmetry about $x=0$ plane need to be broken.  So we consider
that the z direction velocities in the three regions are unequal,
with $v_z^{\rm\uppercase\expandafter{\romannumeral1}}=v_L$,
$v_z^{\rm\uppercase\expandafter{\romannumeral2}}=v$, and
$v_z^{\rm\uppercase\expandafter{\romannumeral3}}=v_R$. For
simplicity, the x and y directions velocities are still with $v_{x/y}^{\rm
\uppercase\expandafter{\romannumeral1}}=v_{x/y}^{\rm\uppercase\expandafter{\romannumeral2}}=
v_{x/y}^{\rm\uppercase\expandafter{\romannumeral3}}=v$.
Figure S1 shows the anomalous velocities induced by the GH shift and IF shift. The anomalous velocities are calculated with the same parameters as in the section ``\textit{Anomalous velocities induced by IF effect}" in the main text. As one can see, the anomalous velocity induced by the GH shift is valley-independent; by contrast, the anomalous velocity induced by the IF shift is valley-dependent.  The green curves and red curves correspond to valley A and valley B, respectively. The thick curves and dashed curves correspond to the anomalous velocities in y-direction and z-direction, respectively. 

Given the incident direction $(\theta, \phi)$ of the incident wave packet, we can calculate the final position $(x, y, z)$ at bottom of the region \uppercase\expandafter{\romannumeral2}. The wave packet needs time $T_0=h/v_{n,z}$ to reach to the bottom of the region \uppercase\expandafter{\romannumeral2} without considering anomalous velocity. However, if we take the anomalous velocity into account, the wave packet needs time $T_1=h/v_{t,z}$ to reach to the bottom of the region \uppercase\expandafter{\romannumeral2}. Thus the anomalous velocities induced position shift is
\begin{eqnarray}\label{eq:s81}
\Delta L=v_{t,y}\times T_1-v_{n,y}\times T_0.
\end{eqnarray}
Put the total velocities of the valley A and valley B into Eq.(\ref{eq:s81}), we can get the final position shifts for the valley A and valley B, respectively [see Fig. 3(d) in the main text].

\vspace{5mm}

\begin{center}
\textbf{${\rm\uppercase\expandafter{\romannumeral4}}$. Calculations of valley density distribution induced by valley-dependent anomalous velocities.}
\end{center}

\begin{figure*}[!htb]
\centering
\includegraphics[height=6cm, width=16cm, angle=0]{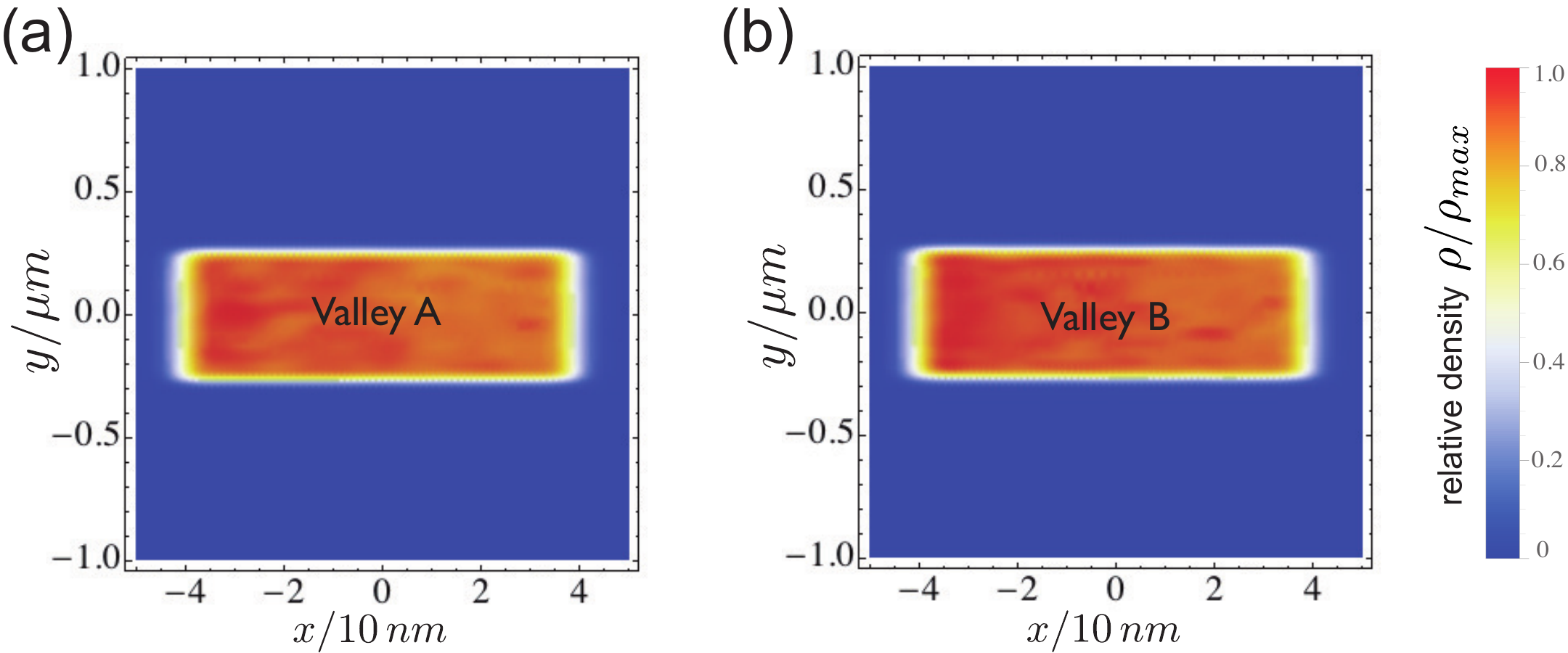}
\caption{{Valley pattern before considering the IF shift.}
The figures (a) and (b) show the valley pattern of valley A and B, respectively, before considering the anomalous velocities induced by the spatial shifts. The color bar on the right represents the relative density of Weyl particles. Here, the relative density has the same meaning as that in the main text. \label{fig:S2}}
\end{figure*}

Since the incident wave packets are likely to be reflected repeatedly at two interfaces, one only needs to consider the total reflected cases since the partially reflected wave packets would eventually diminish after being reflected multiple times.
A collimated incident wave packet usually has certain angle range, thus,
we assume $\theta\in[\theta_c-\delta/2,\theta_c+\delta/2]$ and $\phi\in [\phi_c-\delta/2, \phi_c+\delta/2]$, where $\theta_c$ and $\phi_c$ are the angles of the center of the collimated incident wave packet and
the $\delta$ is the angle spread range.
In the numerical calculations, we take $\theta_c=32.5\degree$ and $\phi_c=90\degree$.
In this case, the anomalous velocity $v_{a,y}$ induced by the GH effect is negative, and the anomalous velocity induced by the IF shift is relatively large. Hence, the total anomalous velocity has a remarkable valley-dependent effect.

For a wave packet of Weyl fermions propagating in the  Region \uppercase\expandafter{\romannumeral2}, the total velocities in  $y$, $z$ directions are $v_{t,y}$ and $v_{t,z}$, respectively. Due to the valley-dependent anomalous velocities, the total velocities are also valley-dependent, which can lead to macroscopic shifts of valleys.  Because of the reflection at the two interfaces (Fig. 3(B)), the  velocity in x direction $v_x^{\uppercase\expandafter{\romannumeral2}}$ in the Region \uppercase\expandafter{\romannumeral2} is a periodic function of $\Delta t$, i.e., $v_x^{\uppercase\expandafter{\romannumeral2}}(t)=v_x^{\uppercase\expandafter{\romannumeral2}}(t+2 \Delta t)$. In the first period, the velocity $v_x^{\uppercase\expandafter{\romannumeral2}}(t)$ reads
\begin{equation}\label{eq:s8}
v_x^{\uppercase\expandafter{\romannumeral2}}(t)=\left\{\begin{array}{cc}
v_x   & t \in [0, \frac{\Delta t}{2})\cup[\frac{3\Delta t}{2},2\Delta t]\\
-v_x  & t \in [\frac{\Delta t}{2},\frac{3\Delta t}{2})\end{array}\right..
\end{equation}
The height of this structure is $h$, thus the wave packet needs time $t_0=h/v_{t,z}$ to reach to the bottom of the Region \uppercase\expandafter{\romannumeral2}. Hence, one can get the final position coordinate $(x, y, z)$ of the wave packet,  where $x=\int_0^{t_0}dt\,{v^{\uppercase\expandafter{\romannumeral2}}_x}(t)$,  $y=\int_0^{t_0}dt\, v_{t,y}$, and $z=-h$.
As one can see, a single incident direction $(\theta, \phi)$ of the wave packet corresponds to a single point $(x, y, z)$ at the bottom of  Region \uppercase\expandafter{\romannumeral2}.  We further assume that the injector supplies an equal strength of incident beams in the angle range $\theta\in [32.5\degree-\delta/2,32.5\degree+\delta/2]$ and $\phi\in [90\degree-\delta/2, 90\degree+\delta/2]$. Thus, there will be a pattern (density distribution)  at the bottom of  Region \uppercase\expandafter{\romannumeral2} [see Fig. 4(d) in the main text]. The Figure S2 (a) and (b) show the density distribution of valley A and valley B, respectively, before considering the IF shift. From the Figure S1(a) and (b) we can find that the density distribution of valley A and valley B are located at the same place. Therefore, the Weyl fermions from valley A and valley B are mixed together. However, if the anomalous velocity is taken into account, the distribution of valley A and valley B will shift to opposite direction leading to the valley separation [as shown in Fig. 4(a), (b) in the main text].

\begin{center}
\textbf{Supplementary References}
\end{center}

\end{widetext}


\begin{thebibliography}{}
\bibitem{footnote2}
Note that the velocity parameters are closely related to the hopping parameters, which can be tuned by strain on the bulk materials. See reference [37] for details.
\end{thebibliography}
\end{document}